\documentclass[10pt,journal,compsoc]{IEEEtran}
\ifCLASSOPTIONcompsoc
  \usepackage[nocompress]{cite}
\else
  \usepackage{cite}
\fi
\ifCLASSINFOpdf
\else
\fi
\usepackage{url}

\usepackage{capt-of}
\usepackage{tabularx}

\usepackage{booktabs} 
\usepackage{times}

\usepackage{datetime}
\PassOptionsToPackage{hyphens}{url}
\usepackage{hyperref}
\usepackage{url}

\usepackage{subfigure}
\usepackage{graphicx}

\usepackage{listings}
\usepackage{color}

\usepackage[utf8]{inputenc}
\usepackage[english]{babel}
\usepackage{multirow}

\usepackage{algorithm}
\usepackage{algpseudocode}
\algdef{SE}{Case}{EndCase}[1]{\textbf{case} \(\mbox{#1}\) \textbf{:}}{\textbf{end}}%
\algdef{SE}[If]{NoThenIf}{EndIf}[1]{\algorithmicif\ #1}{\algorithmicend\ \algorithmicif}%
\algtext*{EndCase}
\algtext*{EndFor}
\algtext*{EndIf}

\usepackage[T1]{fontenc}
\usepackage{textcomp}
\usepackage[none]{hyphenat}

\usepackage[hyphenbreaks]{breakurl}

\lstdefinestyle{C++}{
language=[ISO]{C++},
basicstyle=\ttfamily\scriptsize
}

\newcounter{myenumi}
\renewcommand{\themyenumi}{(\arabic{myenumi})}
\newenvironment{myenumerate}{%
\setlength{\parindent}{0pt}
\setcounter{myenumi}{0}
\renewcommand{\item}{
\par
\vspace{0.1cm}
\refstepcounter{myenumi}
\makebox[1.7em][c]{\themyenumi}
}
}{
\par
\noindent
\ignorespacesafterend
}

\begin{document}
%
\title{StackVault: Protection from Untrusted Functions}
%
%
%
%

\author{Qi~Zhang, Zehra~Sura, Ashish~Kundu, Gong~Su, Arun~Iyengar, Ling~Liu
\IEEEcompsocitemizethanks{\IEEEcompsocthanksitem Q. Zhang, Z. Sura, A.Kundu, G. Su, A. Iyengar
are with IBM Thomas J. Watson Research Center, NY, USA.\protect\\
E-mail: q.zhang@ibm.com, \{zsura, akundu, gongsu, aruni\}@us.ibm.com
\IEEEcompsocthanksitem L. Liu is with Georgia Institute of Technology, GA, USA.\protect\\
E-mail: ling.liu@cc.gatech.edu}
}

\IEEEtitleabstractindextext{%
\begin{abstract}
Data exfiltration attacks have led to huge data breaches. Recently, the Equifax attack affected 147M users and a third-party library - Apache Struts - was alleged to be responsible for it. These attacks often exploit the fact that sensitive data are stored unencrypted in process memory and can be accessed by any function executing within the same process, including untrusted third party library functions. This paper presents \emph{StackVault}, a kernel-based system to prevent sensitive stack-based data from being accessed in an unauthorized manner by intra-process functions. Stack-based data includes data on stack as well as data pointed to by pointer variables on stack. {\em StackVault} consists of three components: (1) a set of programming APIs to allow users to specify which data needs to be protected, (2) a kernel module which uses \emph{unforgeable function identities} to reliably carry out the sensitive data protection, and (3) an LLVM compiler extension that enables transparent placement of stack protection operations. The \emph{StackVault} system automatically enforces stack protection through spatial and temporal access monitoring and control over both sensitive stack data and untrusted functions. We implemented \emph{StackVault} and evaluated it using a number of popular real-world applications, including gRPC. The results show that \emph{StackVault} is effective and efficient, incurring only up to 2.4\% runtime overhead.
\end{abstract}

}

\maketitle

\IEEEdisplaynontitleabstractindextext

%
\IEEEpeerreviewmaketitle

\section{Introduction}
One of the central trust assumptions of software systems is that a function can access data in the memory of another function in the same process. By exploiting this trust assumption, malicious functions can carry out  "function-based data access attacks"  to access sensitive data in the memory of other functions, thus compromising data security and enabling a channel for data exfiltration attacks. Such sensitive data can include protected health information (PHI) and sensitive personal information (SPI), which are required to be protected by privacy regulations such as  GDPR~\cite{gdpr} and HIPAA~\cite{hipaa}. 

Function-based data access attacks can be enabled by malicious code or security bugs in functions in the code-base.
Today, the use of third-party libraries and open source code in software services and products is widespread, and this has increased the use of untrusted functions included in application code~\cite{pittenger2016know,pattabiraman2008samurai}.  
Untrusted functions may not have been subject to rigorous in-house software development and testing practices, thus they could contain security bugs.
Such security bugs can be introduced from multiple sources.
First, open source code repositories have been demonstrated to be susceptible to vulnerabilities \cite{githubHack}, and an increased number of compromises have been observed in the recent past~\cite{hackedrepo}. This makes it possible for security bugs to be introduced into open-source code with malicious intent. Second, inadvertent programming errors can also bring in such bugs.  For example, Equifax recently blamed security holes in Apache Struts to be the reason behind the 147 Million user records data breach~\cite{luszcz2018apache,equifax}. In another example, Heartbleed occurred due to a security bug (inadvertent or malicious in nature) in the open source OpenSSL code~\cite{heartbleed, prevent-heartbleed}. Third, the bugs may also be injected by malicious inside members of a software engineering team. Insiders may contribute new code or sabotage existing code in order to implant security bugs that allow access to sensitive data in the memory of other functions, even when there is no need-to-know~\cite{bishop2003computer} such data. It is reported that a Tesla employee had changed parts of the company's manufacturing operating system code and sent "highly sensitive" company data to outside parties \cite{teslaHack}.

 
  There are several works on security of stack memory~\cite{chen2015stackarmor,dang2015performance,abadi2009control, Strackx:2009:BMS:1519144.1519145} as well as heap memory~\cite{novark2010dieharder, kharbutli2006comprehensively, jang2014safedispatch,nikiforakis2013heapsentry, Silvestro2017FreeGuardAF}. They primarily focus on attacks based on corrupting the contents of heap, or exposing/compromising the stack address and corrupting the contents of stack memory. 
  However, to the best of our knowledge, data access isolation between functions 
  in the same process has not been addressed in the literature so far.
  Shreds~\cite{chen2016shreds} and hardware architectures such as  SGX~\cite{costan2016intel} and SecureBlue++~\cite{boivie2012secureblue++} protect from  outside-in memory access by external processes/threats; outside-in threat is outside of a process and goal is to inject/read  data  in the memory of the process. But
    "function-based data access attacks"  enable inside-out threat, where the threat is inside the process and attacker goal is to exfiltrate data outside the process. 
    
  
 
 

In this paper, we address the problem of how to protect stack-based data from function-based data access attacks. Our focus is to protect data on stacks and also the data in memory pointed to by stack variables (we call these \emph{stack-based data} in this paper). Protecting the stack is a different problem from protecting the heap because stack frames are allocated in a systematic and managed manner and are transient by design. This makes it possible to devise a more efficient solution for protecting stack-based data compared to more general memory protection solutions. Further, in the context of using third-party libraries, the risk of data breaches and exfiltration can be significantly reduced by protecting just the stack-based data.


{\em Our contributions}:
We present the design, implementation and experimental evaluation of our StackVault system that protects stack-based sensitive data from untrusted functions. This system provides the following guarantees: (1) while a sensitive function is executing, the sensitive stack-based data of the function should be preserved in a secure way (protection against corruption), and it  should be protected  from being read with malicious intent 
by untrusted functions (protection against data leakage); (2) when an  untrusted function returns to its caller, the contents of its stackframe are cleared in order to prevent the "untrusted" contents to remain on stack of the thread/process, which can later be used to carry out/enable attacks such as a multi-stage attack; (3) when an untrusted function returns to a caller that is a sensitive function, the contents of the sensitive function's stackframe are restored; (4) when a sensitive function returns to its caller, the contents of its stackframe are cleared. StackVault is designed to support sequential, nested and recursive function calls. 


{\em How StackVault works}: StackVault introduces a notion of "unforgeable function identity" and a kernel based mechanism to protect the sensitive data on the stack. It consists of three components: a set of programming APIs for users to specify what needs to be protected in the source code, a  compiler to compile the source code with StackVault protection, and a kernel module with StackVault related system calls to carry out data protection. 

Our design of StackVault encourages a practice for more efficient and cleaner secure programming; a developer should manage the sensitive data on the stack instead of scattering the sensitive data across memory by using both stacks and heaps, which increases the attack surface. Using only heaps is not always possible because function calls rely on stacks. 


\section{Attack and Threat Model}
\begin{figure*}[htb]
\centering
\subfigure[\small Sequential: f2() is invoked after f1()]{ \includegraphics[width=2.7in, height=1.6in]{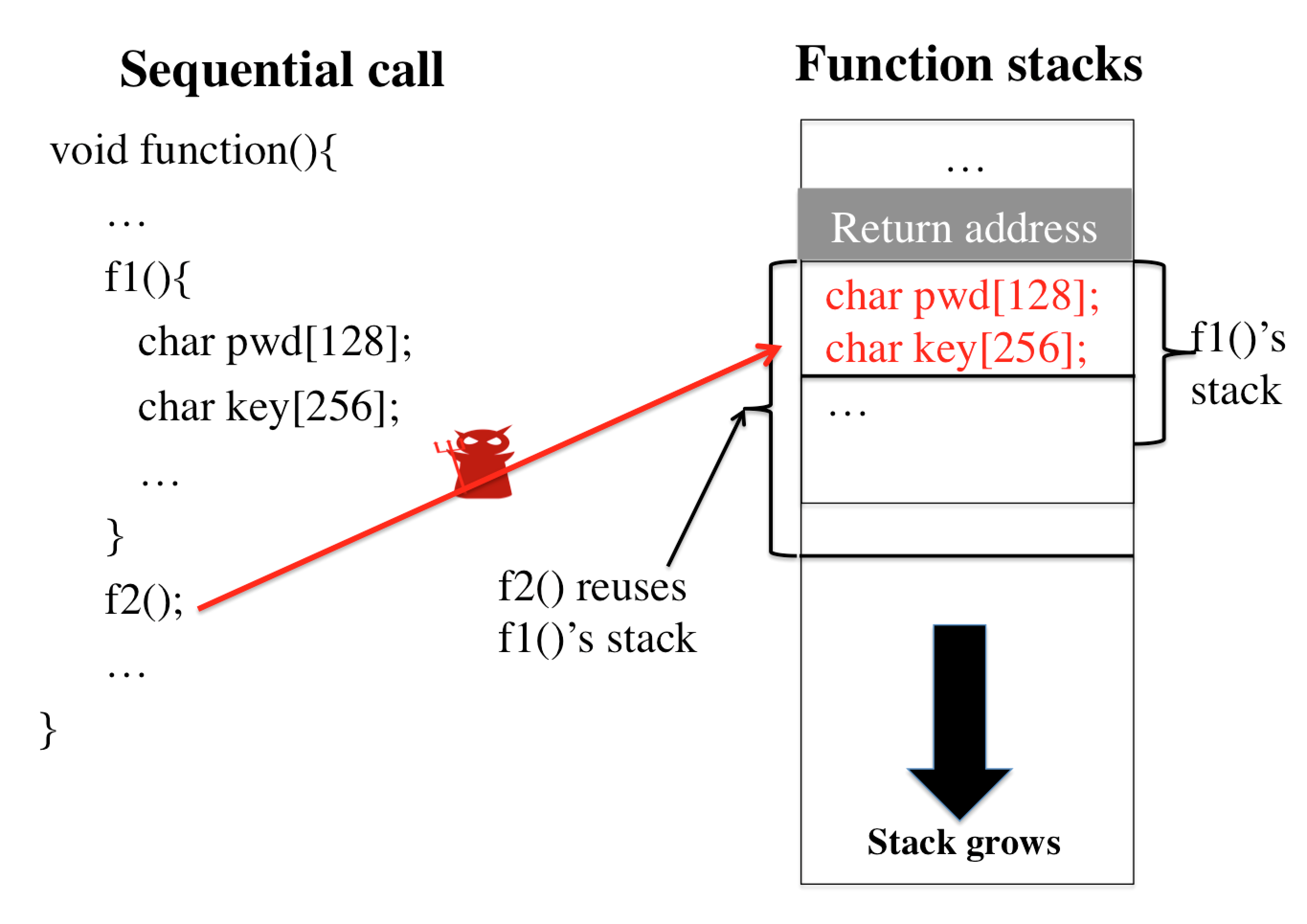}}
\subfigure[\small Nested: f2() is invoked within f1()]{ \includegraphics[width=2.7in, height=1.6in]{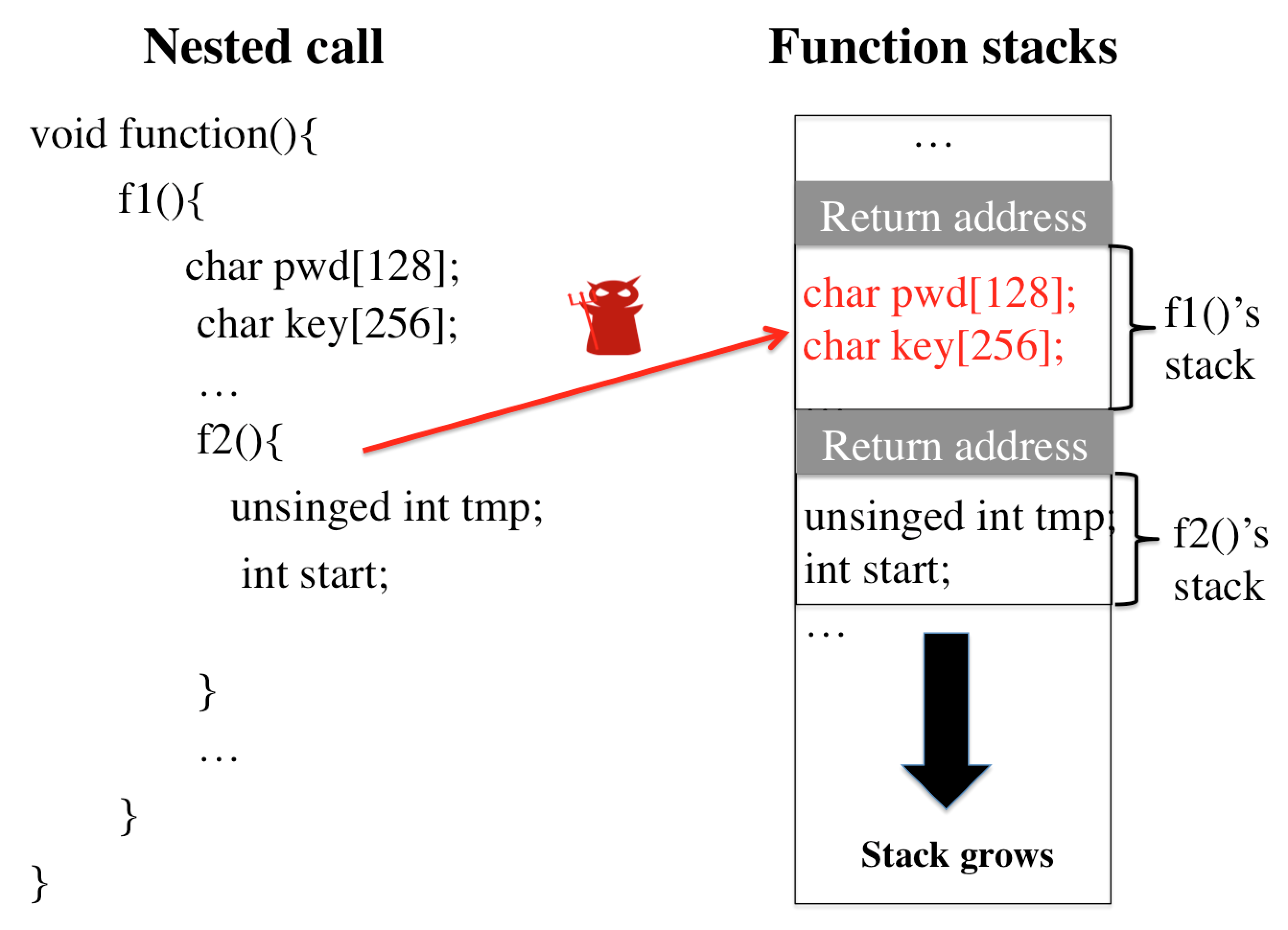}}
\caption{\label{two_cases} {\small Function-based data access attacks in sequential or a nested function call.}}
\end{figure*}

\emph{Function-based Data Access Attacks}.
The underlying assumption in today's programming models "trusts" that a function shall enforce a "need-to-know" policy whenever it accesses the data in the stack and heap of other functions in the code. However, with proliferation of unauthorized/un-regulated open source code and libraries, legacy software, as well as insider threats with access to enterprise software supply chain, such an implicit trust assumption may no longer be valid.

In a function-based data access attack, untrusted functions can access and steal  sensitive data  from the stack-frames and in the heap  of other functions that have been called earlier during the program execution. The vulnerability exploited by attackers is -- any function can access the stack or the heap of another function in the same process. Such access is allowed in almost all processes and is an implicit trust assumption in software systems.




{\em Sensitive and Untrusted Functions}.
A function in a program can be of the following two kinds --  "untrusted function" and "sensitive function"\footnote{Trusted-not-sensitive function: for the sake of completeness, there is another kind of function that is trusted but does not contain/process/produce sensitive data. Such functions are not directly relevant to  this paper, so we do not discuss these kind of functions.}.
\begin{itemize}
     \item Untrusted function: an untrusted function can be any function whose behavior is not fully controlled by a user. Third party libraries are one category of untrusted functions, since they are not fully written in-house and can be compromised by being downloaded from a fake website. In another example, when multiple companies are contributing code to the same project, a function developed by one company can be an untrusted function for another company that is using the function.
     \item Sensitive function: a function that is fully developed in-house and contains sensitive stack-based data, which can be either a variable or parameter allocated on the stack, or a pointer to heap data.
\end{itemize}



{\em Example}. C/C++ support pointers and pointer arithmetic that can be used by \emph{f2()} against its stack base pointer and get access to \emph{f1()}.   Figure \ref{two_cases} displays the stack layout of two functions \emph{f1()} and \emph{f2()} when they are invoked in either a sequential or a nested manner. We assume that \emph{f1()} is a function that has sensitive data on its stack, while \emph{f2()} is an untrusted function. In the nested case, the untrusted function \emph{f2()} is invoked within \emph{f1()}. Since the sensitive data of \emph{f1()} still resides on its stack during the execution of \emph{f2()}, \emph{f2()} can easily get these data if it is compromised. In the sequential case, \emph{f1()} allocates two pieces of sensitive data $-$ password and key  $-$ on its stack, but does not clear them before it returns. Thus, when the untrusted function \emph{f2()} starts to run, it will be able to access these sensitive data on the stack of \emph{f1()}. 

 In our solution, we assume the OS kernel is trusted and has not been compromised. Note that if an attacker has compromised the kernel,  {StackVault} may need to be implemented using a hardware root of trust or as a complete hardware system (which is out of the scope for this paper).

\section{System Design} \label{design}

A system providing StackVault protection guarantees that all previously allocated memory that has been specified as sensitive will be inaccessible in the scope of an untrusted function. This goal is achieved by assuming a trusted kernel and appropriately hiding sensitive data in kernel buffers during execution of untrusted functions.

\begin{figure}[htp]
\centering
\includegraphics[width=3.3in, height=1.5in]{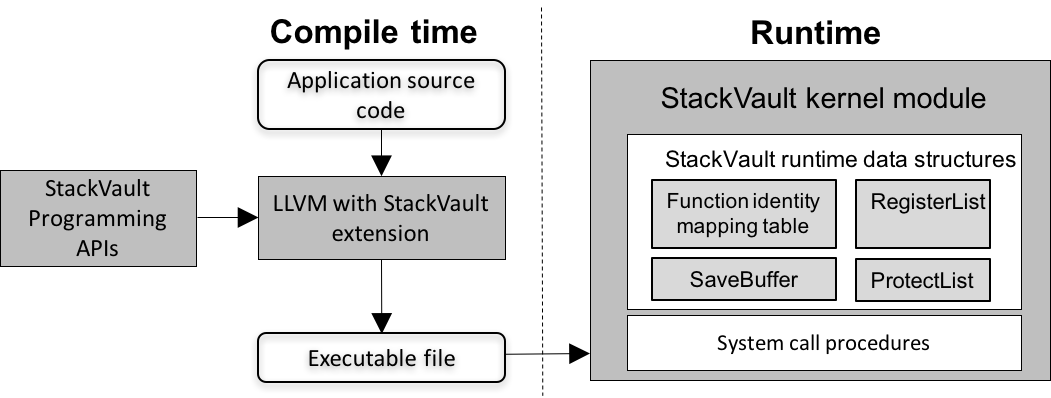}
\caption{StackVault system overview}
\label{system_overview}
\end{figure}

Figure \ref{system_overview} provides an overview of StackVault, which consists of three main components: a StackVault programming API, a Linux kernel module, and an LLVM C/C++ compiler extension. To protect an application with StackVault, programmers first use the programming API to denote the sensitive and untrusted functions in the application source code. 
Next, the source code is compiled using a modified LLVM compiler, which automatically injects StackVault system calls into the executable file. 
Then, during execution of the application, these system calls invoke the StackVault kernel module to protect sensitive stack-based data. 
The following sections describe in detail each of the three StackVault components.

\subsection{StackVault Programming API}
\label{program-api}
In our work, we rely on developers to explicitly specify 
untrusted functions 
and sensitive functions with stack-based data to protect. 
It is possible for an automatic or semi-automatic framework to be developed for this purpose, 
e.g. by using formal methods, analysis techniques, or heuristic rules. 
Such a framework is complementary to our work
and can leverage our system by 
generating code with StackVault APIs appropriately inserted. 
The StackVault programming API provides the capability to:
\begin{myenumerate}
\item {\em Specify untrusted functions}: 
\par 
Untrusted functions are specified by 
listing the function prototypes in a supplementary file called {\em UntrustedList}. 
The format is similar to declaring functions in C/C++ header files. 
For an untrusted third-party library, the header file defining the library functions
can be directly used as the supplementary file. 

\item {\em Specify sensitive functions}: 
\par 
Sensitive functions contain sensitive data (or references to sensitive data) 
in parameters or local variables. 
Like untrusted functions, they are specified using a supplementary file called {\em SensitiveList}. 
Alternatively, sensitive functions can be specified by 
an attribute directive inserted directly in the code 
at the point of the function definition, as follows:

\centerline{\footnotesize{\_\_attribute((annotate(“StackVault\_Sensitive”))) \bf{void foo(int x, …)}\{…\}}} 

By default, the entire stack frame of a sensitive function 
is assumed to be a sensitive memory region. 
However, the programmer can use the annotation string
{\em StackVault\_Sensitive\_Finegrained} to indicate that 
the entire stack frame should not be protected.


\item {\em Annotate individual parameters and local variables}: 
\par
For a sensitive function, any of its parameter or local variable declarations 
can be prepended with an attribute directive 
to allow fine-grained protection and access control. 
For example, a single parameter in the stack frame can be specified to be sensitive, as follows:
\centerline{\footnotesize{\_\_attribute((annotate(“StackVault\_Sensitive\_Finegrained”))) void foo(}}
\centerline{\footnotesize{\_\_attribute((annotate(“StackVault\_Sensitive”))) \bf{int x}, …)\{…\}}} 
The attribute string {\em StackVault\_Sensitive} specifies 
the corresponding variable as a sensitive one to be protected 
from being accessed in untrusted code, 
whereas the attribute string {\em StackVault\_NotSensitive} specifies 
the corresponding variable as one that can be freely accessed. 
The attribute string 
{\em StackVault\_WriteSensitive} specifies a variable to be read-only 
in untrusted code, which ensures that  
the value stored in the variable will not be overwritten
by untrusted function calls.

\item {\em Annotate pointer type parameters and local variables}:
\par 
For a sensitive variable of pointer type, 
the address stored in the pointer is protected. 
However, by default, the memory pointed to by the pointer is not protected. 
The annotation string {\em StackVault\_SensitivePointer} is provided 
to indicate that both the pointer and the memory object pointed to are sensitive.
Likewise, the annotation string {\em StackVault\_WriteSensitivePointer} specifies 
that both the pointer and the memory object are read-only in untrusted code.

\par 
In some cases, the size of the memory object can be automatically determined 
based on the type of the pointer. 
Otherwise, the user has to explicitly provide the size, 
by appending it to the annotation string in the format 
{\em StackVault\_SensitivePointer\_x}, 
where x is the size of the memory object in bytes.
This is useful for handling void pointers and pointer-based arrays.

\par
In our work, we handle protection for only one-level pointer-based memory and assume that the memory objects are contiguous blocks of memory.
In general, memory objects can be recursive pointer-based data structures, 
and there is ongoing work in the community to investigate how shape descriptors 
for arbitrary data structures can be efficiently specified\cite{WOLFE201815}.
However, for all the applications we considered in the context of StackVault, assuming one-level contiguous pointer objects is sufficient to enable effective protection.
\end{myenumerate}

Figure~\ref{fig:examplecode_api} illustrates the use of the StackVault API 
with an example code snippet. 
Figure~\ref{fig:examplecode_api}(a) shows the original code where 
{\em pwdgenerator} is a sensitive function, since it has sensitive data 
such as {\em passwd} and {\em id} allocated on its stack.
{\em lib\_func} is a third-party untrusted function which needs to 
access the {\em age} variable on the stack frame of the sensitive function, 
but must not be allowed to access any other data in that stack frame.
Figure~\ref{fig:examplecode_api}(b) shows the modified code that uses the 
StackVault programming API to enable this protection.

We defined the programming API such that it provides features necessary to express all possible cases of protection requirements.
However, most of cases in practice only need 
to specify untrusted functions and sensitive functions, 
and this can be done without any changes in the source code.
  
\begin{figure*}[ht!]
\centering
\newsavebox\mybox
  \begin{lrbox}{\mybox}
  \begin{minipage}{0.44\textwidth}
  \begin{lstlisting}[firstnumber=1]%[frame=none,numbers=none]
  ...
/*pwdgenerator is a sensitive function*/
pwdgenerator(){ 
  /*sensitive data on stack*/
  char passwd[256];
  /*stack pointer pointing to sensitive memory*/
  char *id;
  /*not sensitive data; accessed by lib_func*/
  int age; 
  ...
  id = malloc(len);
  ...
  
  /*lib_func is an untrusted function*/
  lib_func(&age);
  ...
}
  \end{lstlisting}
  \end{minipage}
  \end{lrbox}
  \subfigure[Original code]{\usebox\mybox}
  \label{sysdesign:origcode}
  \quad
  \begin{lrbox}{\mybox}
  \begin{minipage}{0.54\textwidth}
  \begin{lstlisting}[firstnumber=1]%[frame=none,numbers=none]
  ...
__attribute((annotate('StackVault_Sensitive')))
pwdgenerator(){ 
  
  char passwd[256];
  __attribute((annotate('StackVault_SensitivePointer_len')))
  char *id;
  __attribute((annotate('StackVault_NotSensitive')))
  int age; 
  ...
  id = malloc(len);
  ...
  
  /* add prototype of lib_func in UntrustedList */
  lib_func(&age);
  ...
}  
  \end{lstlisting}
  \end{minipage}
  \end{lrbox}
  \subfigure[Code with \emph{StackVault} API]{\usebox\mybox}
  \label{sysdesign:codeAPI}
  \quad
\caption{Example code using \emph{StackVault} protection}
\label{fig:examplecode_api}
\end{figure*}

\subsection{StackVault Kernel Module}
\label{kernelruntime}


The StackVault kernel module provides a set of system calls to specify \emph{what} stack-based memory to protect and to determine \emph{when} to protect. Protection is guaranteed for an executable binary generated with these system calls inserted at appropriate locations. On loading an application binary, the kernel scans the text section (using an ELF parser) to record the contiguous instruction address span for each named function in the code. These instruction address spans then serve as \emph{unforgeable function identities}; i.e. the kernel can ascertain whether a given program counter (PC) register value during execution corresponds to the code of a specific function or not, without the possibility of any function masquerading as another. These function identities are reliable due to two reasons: first, the text sections in executable binaries are read-only and cannot be dynamically modified, 
and second, the value of PC registers cannot be modified by malicious user-level code.

The kernel runtime maintains four data structures: a \emph{Function identity mapping table}, a \emph{RegisterList} to record what is to be protected, a \emph{ProtectList} to record the sequence of the protected calls, and a \emph{SaveBuffer} to temporarily save sensitive data.
These data structures are only available to kernel code, 
and cannot be accessed from user space.

The following system calls are used to specify what to protect:
\begin{myenumerate}
\item {\bf register\_stack(bool all)}:
\par
This call is to be invoked at the very beginning of each 
sensitive function. The boolean parameter determines whether the entire stack frame 
is to be protected.
When invoked, this call automatically determines the 
identity of the calling function based on the current PC register value.
It also determines the bounds of the stack frame for the calling function 
based on the calling convention 
(using {\em rbp} and {\em rsp} register values on x86 architecture).
It then creates a new entry in the {\em RegisterList},  
and records the function identity, address bounds of the stack frame, 
and the boolean parameter value passed in.

\item {\bf register\_memory(char *base, unsigned long len, bool readOnly)}:
\par
This call is used to specify fine-grained regions of the stack frame 
that are sensitive and to be protected. 
It is also used to protect the memory object pointed to by 
a sensitive pointer variable on the stack.
When invoked, this call first verifies that the identity of the calling function 
is the same as the function identity corresponding to the last 
register\_stack call recorded in the {\em RegisterList}. 
If it does not match, an exception is flagged by the kernel, since a different identity indicates an illegal invocation.
Otherwise, the kernel creates a new entry in the {\em RegisterList},  
and records the parameters passed in, which are the base address,  
length in bytes of the memory region to be protected, 
and the type of protection (read-only access, 
or full protection).

\item {\bf register\_memory\_exception(char *base, unsigned long len, bool readOnly)}:
\par
This call is used to specify fine-grained regions of a sensitive stack frame 
that must be accessible from within an untrusted function.
It is particularly useful when the address of a variable 
on the stack of a sensitive function 
is passed as a parameter to the untrusted function.
As in the previous case, this call first verifies that 
the identity of the calling function is the same as 
the last register\_stack call recorded in the {\em RegisterList}.
It also verifies that the memory region defined by \emph{base} and \emph{len} 
falls within the current stack frame.
If so, it creates a new entry in the {\em RegisterList},  
and records the parameters passed in. 


\item {\bf unregister\_stack(void)}:
\par
This call is to be invoked right before a sensitive function returns.
As in the previous case, this call first verifies that 
the identity of the calling function is the same as 
the last register\_stack call recorded in the {\em RegisterList}.
If so, it removes the {\em RegisterList} entry of the last 
register\_stack call and all subsequent entries to it as well.
It also clears the stack frame of the calling function, 
which ensures that untrusted functions invoked later in execution will not 
be able to access any leftover sensitive stack data.

\end{myenumerate}

The following system calls are used to determine when to protect:
\begin{myenumerate}
\item {\bf start\_protect(void)}:
\par
This call is to be invoked right before an untrusted function call. 
It enables protection for all registered sensitive memory regions 
by copying the sensitive data to kernel buffers. 
It also clears the sensitive memory regions 
(except when the memory region is allowed read-only access) 
to ensure that sensitive data is not leaked to untrusted functions.
This call also determines the identity of the calling function 
and creates a new entry in the {\em ProtectList} 
recording the function identity and the index of the 
first free entry in the {\em RegisterList}.
\par
Note that in the case of nested calls to start\_protect, 
protection will already be in effect for sensitive data 
registered prior to the previous start\_protect call, 
and this invocation of start\_protect needs to process 
only the data newly registered since then. 
These {\em RegisterList} entries are processed in order, and 
Algorithm~\ref{algo:startprotect} gives details of how this is done.

\item {\bf stop\_protect(void)}:
\par
This call is to be invoked right after an untrusted function call returns. 
On invocation, it first determines the identity of the calling function,
and verifies that it is the same as the function identity 
recorded in the last entry in the {\em ProtectList}. 
If the identities do not match, an exception is flagged.
Then it verifies that the index of the first free entry 
in the {\em RegisterList} matches the index recorded in the last entry 
in the {\em ProtectList}.
If it does not match, then there are improper (possibly malicious)
entries in the {\em RegisterList} added during execution of untrusted code, 
and an exception is flagged.
\par
Otherwise, the kernel removes the last {\em ProtectList} entry.
Then it uses the $SaveBuffer$ 
to restore data for all sensitive memory regions. 
This ensures that the application continues executing correctly 
and remains unaffected by any corrupt data 
that may have been written into sensitive areas by untrusted code.
The restoration of stack-based data is done by processing 
{\em RegisterList} entries in order, 
analogous to the processing for start\_protect calls.
This procedure assumes that for a given sensitive function, 
there is no overlap between the memory objects referred to by 
register\_memory and register\_memory\_exception calls.
This constraint is automatically satisfied when the 
StackVault programming API and compiler are used to generate the executable code.
\par
Algorithm~\ref{algo:stopprotect} gives details of how 
the stack frame restoration is done.
Note that the pseudocode shown here gives the logic 
of the algorithm; the actual implementation is optimized 
for the common case.
Also note that data written to {\em SaveBuffer} 
is read back exactly once, 
and buffer management can free up the corresponding space on a read.

\end{myenumerate}


\begin{algorithm}[ht]
\caption{Algorithm for processing {\em start\_protect} calls}
\label{algo:startprotect}
\footnotesize
\begin{algorithmic}[1]
\If{$ProtectList$ is empty}
    \State $start \gets 0$
\Else
    \State $start \gets$ index stored in last $ProtectList$ entry
\EndIf
\State $end \gets$ index of last entry in $RegisterList$
\For{$i \gets start, end$}
   \If{$RegisterList$[$i$] is $register\_stack$}
       \If {$all$ is true}
           \State copy stack frame to $SaveBuffer$
       \EndIf
   \Else
       \State copy memory object to $SaveBuffer$
   \EndIf
\EndFor
\For{$i \gets start, end$}
   \Case{$RegisterList$[$i$] is $register\_stack$}
       \If {$all$ is true}
           \State clear stack frame
       \EndIf
   \EndCase{}
   \Case{$RegisterList$[$i$] is $register\_memory$}
       \If{$readOnly$  is false}
           \State clear memory object
       \EndIf
   \EndCase{}
   \Case{$RegisterList$[$i$] is $register\_memory\_exception$}
       \State copy back memory object from $SaveBuffer$
   \EndCase{}
\EndFor
\end{algorithmic}
\end{algorithm}

\begin{algorithm}[ht]
\caption{Algorithm for processing {\em stop\_protect} calls}
\label{algo:stopprotect}
\footnotesize
\begin{algorithmic}[1]
\State {\bf var} $TempStack$ {\footnotesize \em{/*temp buffer used to assemble stack data*/}}
\State {\bf var} $StackAddr \gets$ NULL
\If{$ProtectList$ is empty}
    \State $start \gets 0$
\Else
    \State $start \gets$ index stored in last $ProtectList$ entry
\EndIf
\State $end \gets$ index of last entry in $RegisterList$
\For{$i \gets start, end$}
   \Case{$RegisterList$[$i$] is $register\_stack$}
       \If {$StackAddr$ is not NULL}
           \State /* copy previous stack frame to orig location*/
           \State [$StackAddr$] $\gets TempStack$ 
       \EndIf
       \State $StackAddr \gets RegisterList$[$i$] stack address
       \State /* copy current stack frame to temp location*/
       \If {$all$ is true}
           \State [$TempStack$] $\gets$ from $SaveBuffer$
       \Else
           \State [$TempStack$] $\gets$ [$StackAddr$]
       \EndIf
   \EndCase{}
   \Case{$RegisterList$[$i$] is $register\_memory$}
       \If{memory object is on stack frame}
           \State copy from $SaveBuffer$ to $TempStack$
       \Else
           \State copy from $SaveBuffer$ to its original location
       \EndIf
   \EndCase{}
   \Case{$RegisterList$[$i$] is $register\_memory\_exception$}
       \If{memory object is on stack frame}
           \State copy from original location to $TempStack$ 
       \EndIf
   \EndCase{}
\EndFor
\If {$StackAddr$ is not NULL}
    \State [$StackAddr$] $\gets TempStack$
\EndIf
\end{algorithmic}
\end{algorithm}

Figure~\ref{fig:examplecode_syscalls} shows the example of Figure~\ref{fig:examplecode_api} 
with appropriate StackVault system calls inserted.

\begin{figure}[h!]
\centering
  \begin{lrbox}{\mybox}
  \begin{minipage}{0.5\textwidth}
  \begin{lstlisting}[firstnumber=1]%[frame=none,numbers=none]
  ...
pwdgenerator(){ 
  register_stack(all=True);
  char passwd[256];
  char *id;
  int age; 
  ...
  id = malloc(len);
  ...
  register_memory(id, len, False);
  register_memory_exception(&age, sizeof(int), False);
  start_protect();
  lib_func(&age);
  stop_protect();
  ...
  unregister_stack();
}  
  \end{lstlisting}
  \end{minipage}
  \end{lrbox}
  \subfigure{\usebox\mybox}
\caption{Example code with \emph{StackVault} system calls}
\label{fig:examplecode_syscalls}
\end{figure}

\paragraph{System Call Invocation in Assembly}

Although StackVault kernel runtime calls are designed as system calls, 
they need to be invoked 
using inline assembly (\_\_asm\_\_ in C/C++) 
and are identified by the system call number,
instead of being invoked directly using the system call names. 
This is because whenever a StackVault system call is invoked, 
it refers to the instruction pointer (saved PC register value) 
to determine the identity of the function that issued the invocation, 
so that illegal invocations from untrusted functions can be detected. 
However, when system calls are invoked by their names, 
the instruction pointer obtained within the system call 
will always point to the same address in kernel code, 
regardless of the function invoking the system call. 
This is because the name of the system call is a wrapper function, 
and the real system call is invoked inside this wrapper. 
Therefore, if the system call is invoked by its name, 
the StackVault system call will not be able to determine the unforgeable identity of the function that invokes the system call. 

\newcolumntype{C}{>{\centering\arraybackslash}X}
\begin{table*}
  \caption{Mapping from StackVault API to system calls}
  \label{tab:apimap}
  \begin{tabularx}{\textwidth}{@{}c*{10}{C}c@{}}
    \hline
    \textbf{API} & \textbf{System Call(s)} & \textbf{Parameter} & \textbf{Comment} \\
    \hline
    Untrusted function & start\_protect {\em and} stop\_protect & & before {\em and} after call \\
    \hline
    Sensitive function & register\_stack {\em and} unregister\_stack & all=True & begin {\em and} end of function \ \ \\
    \hline
    Sensitive\_Finegrained function & register\_stack {\em and} unregister\_stack & all=False & begin {\em and} end of function\\
    \hline
    Sensitive var & register\_memory & readOnly=False &  only when all=False\\
    \hline
    NotSensitive var & register\_memory\_exception & readOnly=False &  only when all=True \\
    \hline
    WriteSensitive var & register\_memory & readOnly=True &  only when all=False \\
    \hline
    WriteSensitive var & register\_memory\_exception & readOnly=True &  only when all=True\\
    \hline
    SensitivePointer var\textbf{$^*$} & register\_memory & readOnly=False & for pointee object\\
    \hline
    WriteSensitivePointer var\textbf{$^*$} & register\_memory &  readOnly=True \ \ \ & for pointee object\\
    \hline
  \end{tabularx}
  \hspace{3cm}
  \footnotesize{\textbf{$^*$}The table row describes handling for the pointee object only. The pointer var is also handled in this API, using the same logic as for non-pointer variables.}
\end{table*}

\subsection{StackVault Compiler Extension}
\label{compiler}
The StackVault compiler automatically transforms code annotated with the StackVault programming API 
to executable binaries with appropriate StackVault system calls inserted.
We modified the LLVM compiler release version 6.0.0 to implement this functionality.
This compiler provides new command-line options: 
-qstackvault to enable the StackVault system, 
and -qstackvault\_dir to specify the directory path containing the 
{\em SensitiveList} and {\em UntrustedList} files. 
When invoked, the compiler front-end processes these files and registers the 
identities of sensitive and untrusted functions.

Then, the code generation phase incorporates the following changes for StackVault processing:
\begin{enumerate}
\item Function calls: If the function being invoked matches the prototype of an untrusted function, 
then the compiler inserts a start\_protect system call after the code setting up the function call 
(i.e.~after the callee stack frame has been allocated and argument values have been copied, 
but before the jump to the callee function code). 
It also inserts a stop\_protect system call in the code generated to process the return, 
right before the jump back to the caller code. 
Further, it checks all arguments of the untrusted function to see if any 
is of pointer type and is assigned the address of a local variable on the caller stack frame. 
For any such argument, it treats the corresponding local variable as 
{\em StackVault\_NotSensitive}, 
and inserts a register\_memory\_exception system call if needed.

\item Function definitions: If the function being generated matches a sensitive function, 
then the compiler inserts a register\_stack system call at the end of the function prologue, 
and an unregister\_stack system call at the beginning of the function epilogue.

\item Local stack variables: Function parameters and local variables typically have space allocated to them on the function stack frame. 
When processing them, the compiler checks the parameter/variable definition to see if there is any StackVault annotation attached. 
If so, it inserts the appropriate register\_memory or register\_memory\_exception system calls at the end of the function prologue.
\end{enumerate}

Table~\ref{tab:apimap} lists all the mappings 
from StackVault programming APIs to StackVault system calls.
For all sensitive and untrusted functions, 
the compiler ensures that the function is not inlined. 
This allows the runtime verification to properly determine the function identity.
Also, the -fno-omit-frame-pointer compiler flag is used 
to make sure the frame pointer register is always set up,
so that the kernel runtime can properly determine the bounds of the 
function stack frame on a register\_stack call. 

StackVault is compiled with link-time optimization enabled, and all StackVault system calls inserted during compilation are inlined.

\section{Security Analysis}
In this section, we show how the StackVault system design makes it robust against adversarial manipulations that attackers may try to use. At runtime, the StackVault kernel module tracks  corresponding pairs of \emph{register\_stack} and \emph{unregister\_stack} calls, and pairs of \emph{start\_protect} and \emph{stop\_protect} calls. It ensures that these calls are invoked in a legal and orderly manner by using runtime verification checks and relying on \emph{unforgeable function identities}. Thus, even when an attacker is familiar with the StackVault system calls, the design prevents evading or misusing StackVault security protection.

First, the attacker may try to {\bf evade} StackVault protection by invoking \emph{stop\_protect} in untrusted functions to illegally restore the protected stack and steal sensitive information from there. StackVault defends against this attack by checking if the function invoking \emph{stop\_protect} is the same as the function that most recently invoked \emph{start\_protect}. If not, then this invocation will raise an exception, and a warning will be given to alert users that their code is under attack. Even if the attacker manages to hijack the program control flow and branch to the execution of the \emph{stop\_protect} call in the appropriate function, any return to untrusted function code will require crossing another \emph{start\_protect} call boundary, and the attack will fail.

The attacker may also try to evade protection by using the \emph{register\_memory\_exception} call in untrusted functions to void protection for specific memory regions. StackVault defends against this attack by checking if the memory region defined by the parameters of a \emph{register\_memory\_exception} falls within the stack frame of the current function. As a result, this call can be used by a function to enable access to its own stack frame only, and it cannot affect any other (possibly sensitive) memory regions.

Second, the attacker may try to {\bf misuse} StackVault protection by invoking \emph{register\_stack} or \emph{register\_memory} in untrusted functions to illegally restrict regular code from accessing memory regions 
that they should have access to.
StackVault detects such misuse on return from untrusted functions, when \emph{stop\_protect} is invoked and the runtime checks if the number of entries 
in the kernel's \emph{RegisterList} is the same as it was when the 
corresponding \emph{start\_protect} was invoked.
This check detects any improper \emph{RegisterList} entries leftover 
from calls made in untrusted functions, and flags a warning to the user.

Note that the function identity checking prevents misuse of 
\emph{unregister\_stack} calls in untrusted functions.
The unforgeable function identities used in StackVault 
cannot be subverted by a user level program 
and are a key component for enforcing protection.
StackVault makes effective use of the PC register (for function identities) 
and stack register (for determining stack frame boundaries)
as defined in the architecture and system-level API.
StackVault also assumes a trusted kernel with privileged access to some memory buffers, 
and a cooperating compiler that does not inline any untrusted functions.

\section{Evaluation}\label{evaluation}
In this section, we first measure the overheads of StackVault by comparing the overall execution time of different applications while running with and without StackVault. Then, we zoom into more detailed StackVault runtime statistics to break down the overheads and investigate various factors that could have different impact on the overheads. Finally, we evaluate how StackVault affects the executable file size and the compilation time .

\subsection{Experimental Setup}
StackVault is evaluated using the following applications: gRPC, xmlstream, fileupload, htmltidy, minizip, and miniunz. In each of the applications, there is sensitive data allocated on the function stack, while the third party library calls are invoked in the same function. Therefore, such library calls could access the sensitive data if not being protected. The following two scenarios are compared in each case:

\begin{list}{}{\setlength{\leftmargin}{1em} \setlength{\topsep}{1ex} \setlength{\itemsep}{1ex}}
\item[$\bullet$] 
\textbf{Native.} The application is compiled with original LLVM; thus it is running without StackVault capabilities.
\item[$\bullet$] 
\textbf{StackVault.} Given a list of sensitive and untrusted functions, the application is compiled with the extended LLVM for StackVault, and it is running with StackVault protections.
\end{list}

\begin{table*}[]
\centering
\small
\caption{Application details}
\label{application_details}
\begin{tabular}{|c|c|c|c|c|c|}
\hline
\textbf{Application} & \textbf{Category}     & \textbf{LoC} & \textbf{sensitive funcs} & \textbf{\# of untrusted funcs} & \textbf{Third Parties} \\ \hline
gRPC              & Remote procedure call benchmark & 6049                      &     \multicolumn{1}{c|}{\begin{tabular}[c]{@{}c@{}}QpsWorker\\RunClientBody\\RunServerBody\end{tabular}}                            & 2                               & gRPC        \\ \hline
libcurl-xmlstream    & File parser           & 103                   & main                              & 14                             & libcurl, expat         \\ \hline
libcurl-fileupload   & File upload           & 41                    & main                              & 6                              & libcurl                \\ \hline
libcurl-htmltidy     & File parser           & 83                    & main                              & 20                             & libcurl, libtidy       \\ \hline
minizip              & File compression tool & 520                   & main                              & 3                              & zlib                   \\ \hline
miniunz              & File compression tool & 671                      & main                               & 3                               & zlib                   \\ \hline

\end{tabular}
\end{table*}

All the experiments are conducted on an Intel Xeon E5640 server with 16 2.67GHz CPUs, 32 GB memory, and 500GB disk storage. Ubuntu 16.04 with Linux kernel 4.4.98 is used as the operating system. All the reported results are averaged over 5 runs. The details of each application used in the experiments are as follows:

\begin{list}{}{\setlength{\leftmargin}{1em} \setlength{\topsep}{1ex} \setlength{\itemsep}{1ex}}
\item[$\bullet$] 
\textbf{gRPC\cite{gRPC}.} gRPC is an open source remote procedure call system initially developed at Google. It is widely used to establish communications among different components in low latency, highly scalable, distributed systems.  In this paper, we use a gRPC benchmark which runs 2 processes called QPS workers that act as the gRPC client and server, as well as one driver process that sets those workers up to run a specific test scenario. The driver sends the configuration parameters to the workers and reports the resulting statistics after the scenario is complete. 
\item[$\bullet$] 
\textbf{xmlstream\cite{xmlstream}.} This application uses the libcurl library to download an XML file from a given URL, and then parses this file via the streaming Expat parser. We run the xmlstream by downloading a 1MB XML file and parsing it.
\item[$\bullet$]  
\textbf{fileupload\cite{fileupload}.} This application uses the libcurl library to upload a file to a given URL. We run the fileupload to upload a 64MB file from one local directory to another local one.
\item[$\bullet$] 
\textbf{htmltidy\cite{htmltidy}.} This application downloads an HTML document using the libcurl library and parses the document using the libtidy library. We run the htmltidy to download the html page from \emph{www.google.com} and parse it.
\item[$\bullet$] 
\textbf{minizip\cite{minizip}.} This application creates a compressed file from a normal file or directory. We run the minizip to compress a folder with 5 image files, which are in total 146MB.
\item[$\bullet$] 
\textbf{miniunz\cite{miniunz}.} This application uncompresses a compressed file. We run the miniunz to uncompresses the files compressed by minizip.
\end{list}

Table \ref{application_details} shows the names of sensitive functions and number of untrusted functions in each application. The untrusted functions are from the \emph{Third Party} libraries listed in the last column of the table. As discussed before, since the source code of such third party libraries are usually not written by the developers of the applications, these developers cannot have full control over the behavior of the functions in such libraries. The sensitive functions are the ones in the applications that have sensitive data on the stack. In practice, it highly depends on the developers of the applications to specifically define which data is sensitive. 
Taking the gRPC benchmark for an example, the functions \emph{RunClientBody()} and \emph{RunServerBody()} allocate sensitive data \emph{ClientArgs} and \emph{ServerArgs} on the stack. At the same time, a gRPC library call \emph{gpr\_log()} is frequently invoked by these two functions. In order to prevent \emph{gpr\_log()} from illegally accessing the sensitive data on the stack, \emph{RunClientBody()} and \emph{RunServerBody()} are selected as sensitive functions, while \emph{gpr\_log()} is selected as one of the untrusted functions. Similarly, \emph{main()} in \emph{libcurl-fileupload} is selected as a sensitive function, since it allocates a sensitive variable \emph{file\_info} on the stack, while also invoking many third party library calls such as \emph{curl\_easy\_getinfo()}, which should not be allowed to access the \emph{file\_info}.

\subsection{Execution time overhead}
In this subsection, we examine StackVault's impact on application performance. 
We instrument applications with \emph{gettimeofday()} to obtain execution times in microseconds for each application in the native case and the StackVault case respectively. 

\begin{table*}[]
\small
\centering
\caption{Number of each StackVault related system calls invoked by the applications}
\label{systemcall_detail}
\begin{tabular}{|c|c|c|c|c|c|c|c|c|}
\hline
\textbf{}  & \textbf{\begin{tabular}[c]{@{}c@{}}register\\ \_stack()\end{tabular}} & \textbf{\begin{tabular}[c]{@{}c@{}}unregister\\ \_stack()\end{tabular}} & \textbf{\begin{tabular}[c]{@{}c@{}}register\_\\ memory()\end{tabular}} & \textbf{\begin{tabular}[c]{@{}c@{}}register\_ \\memory\_\\ exception()\end{tabular}} & \textbf{\begin{tabular}[c]{@{}c@{}}start\_\\ protect()\end{tabular}} & \textbf{\begin{tabular}[c]{@{}c@{}}stop\_\\ protect()\end{tabular}} & \textbf{Total} & \textbf{\begin{tabular}[c]{@{}c@{}}Total stack-based memory\\ copied and cleared \\ in bytes\end{tabular}} \\ \hline
gRPC       & 3                                                                     & 3                                                                       & 0                                                                      & 0                                                                                 & 23                                                                   & 23                                                                  & 52             &26,960                                                                                      \\ \hline
xmlstream  & 1                                                                     & 1                                                                       & 0                                                                      & 1                                                                                 & 112                                                                  & 121                                                                 & 245            &29,200                                                                                        \\ \hline
fileupload & 1                                                                     & 1                                                                       & 0                                                                      & 2                                                                                 & 10                                                                   & 10                                                                  & 24             &5,824                                                                                        \\ \hline
htmltidy   & 1                                                                     & 1                                                                       & 1                                                                      & 147                                                                               & 2,053                                                                 & 2,053                                                                & 4,255           &2,187,680                                                                                      \\ \hline
minizip    & 1                                                                     & 1                                                                       & 0                                                                      & 6                                                                                 & 11                                                                   & 11                                                                  & 30             &32,114                                                                                       \\ \hline
miniunz    & 1                                                                     & 1                                                                       & 0                                                                      & 1                                                                                 & 7                                                                    & 7                                                                   & 17             &16,528                                                                                        \\ \hline
\end{tabular}
\end{table*}

\begin{figure*}[htp]
\centering
\begin{minipage}{.5\textwidth}
  \centering
  \includegraphics[width=3.0in, height=1.5in]{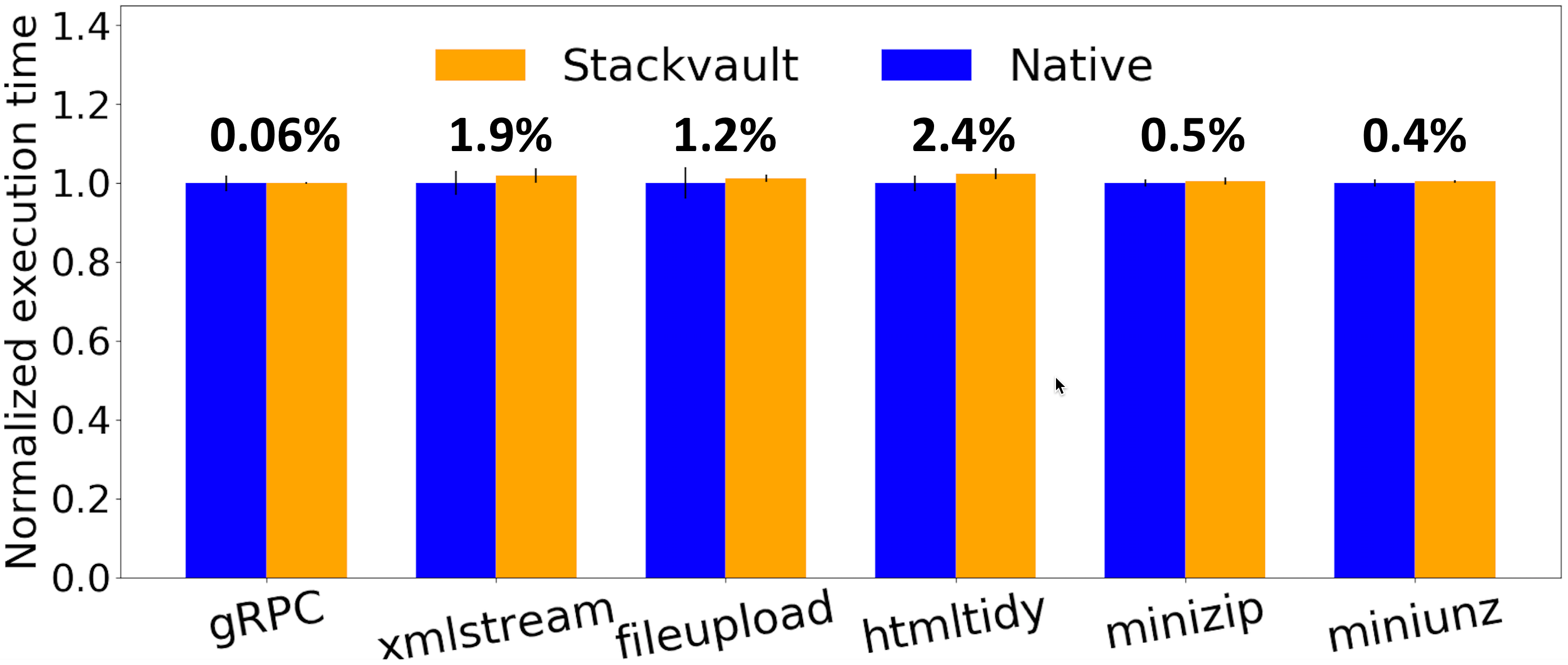}
  \captionof{figure}{Normalized execution time comparison.}
  \label{execution_overhead_total}
\end{minipage}%
\begin{minipage}{.5\textwidth}
  \centering
  \includegraphics[width=3.0in, height=1.5in]{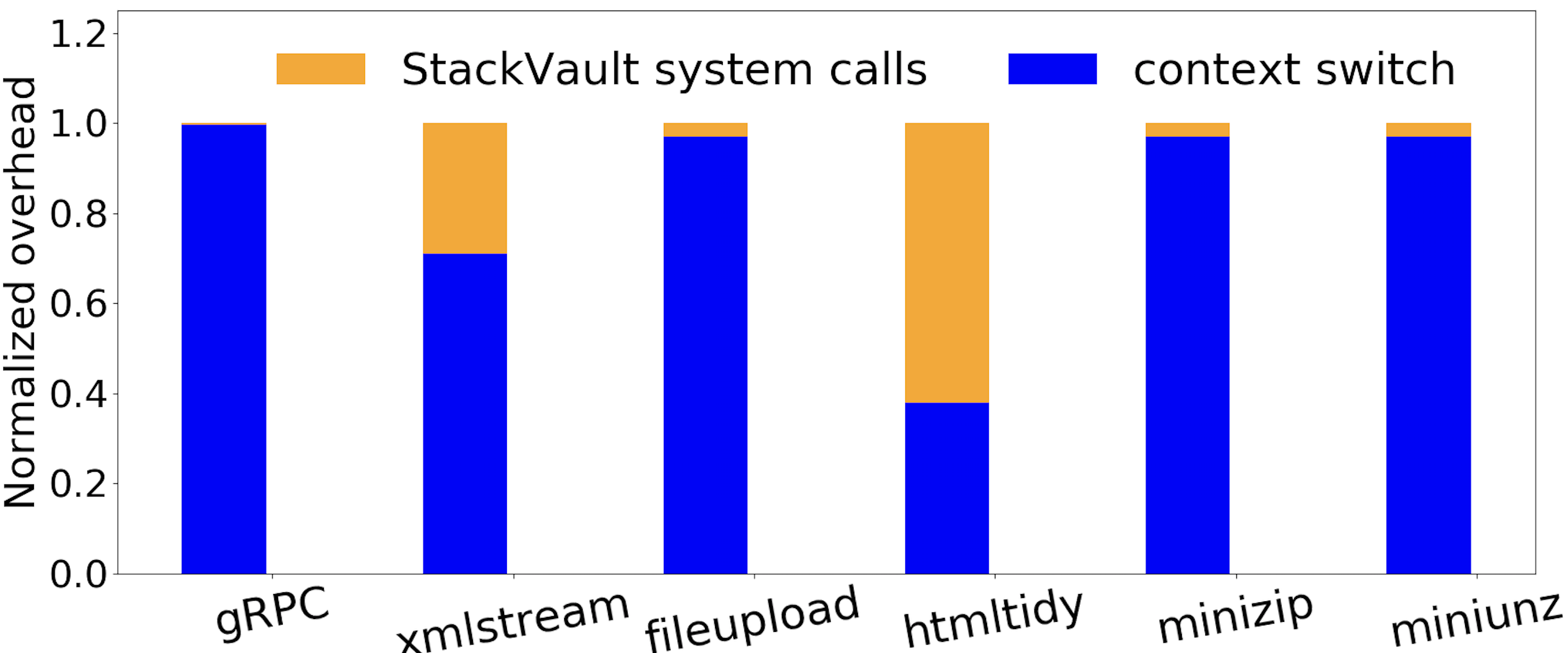}
  \captionof{figure}{Normalized overhead breakdown - context switch v.s. StackVault system call.}
  \label{total_overhead_breakdown}
\end{minipage}
\end{figure*}



The normalized execution time of each application is displayed in Figure \ref{execution_overhead_total}. It shows that the execution time overhead is negligible for all the applications. For example, compared with the Native case, the execution time of fileupload increases 1.2\% in the StackVault case, while that of minizip and miniunz increases 0.5\% and 0.4\% respectively. This indicates that even though additional system calls need to be invoked by the application in order to interact with StackVault, the overhead incurred by such system calls is small.

We zoom into the overhead to understand where it comes from and which part contributes most to such overhead. Basically, the additional execution time consists of two parts: first, since the application needs to invoke the StackVault specific system calls, there is overhead in context switching between user space and kernel space; the second part is the time spent inside each StackVault related system call. Figure \ref{total_overhead_breakdown} displays the normalized time spent on each component. We have two observations. First, for some applications, compared with the time spent inside the StackVault system calls, the context switch dominates the overhead. For instance, in gRPC and fileupload, 99\% and 95\% of the StackVault overhead is due to the context switch. Second, the distribution between the context switch overhead and the StackVault specific system calls overhead varies among different applications. Comparing minizip and xmlstream, although the context switch overhead exceeds the system call overhead for both applications, the context switch occupies 97\% of the total overhead in minizip while it occupies 67\% in htmltidy.

To explain such differences, we measure how frequently StackVault system calls are invoked in each application, and the results are displayed in Table \ref{systemcall_detail}. There are a few interesting observations. First, xmlstream and htmltidy have the most StackVault system calls. This is consistent with the results in Figure \ref{execution_overhead_total}, which show that xmlstream and htmltidy are the two applications that incur the most execution time overhead - 1.9\% and 2.4\% respectively. Second, most of the StackVault system calls invoked by xmlstream and htmltidy are \emph{start\_protect()} and \emph{stop\_protect()}. Recall that in Figure \ref{total_overhead_breakdown}, xmlstream and htmltidy are the two applications that have the highest percentage of StackVault system call overhead; the overheads of \emph{start\_protect()} and \emph{stop\_protect()} are higher than their context switch cost, while that of the other StackVault system calls are less than the context switch overhead. Therefore, the more these two calls are invoked, the higher the system call overhead in terms of percentage. The results also indicate that on average, a context swtich between the user application and a StackVault system call takes 7 microseconds. In addition, the overhead of the system calls themselves depends on many other factors. For example, a larger stack size could bring more overhead since more data needs to be copied between the user stack and the kernel buffer.

\begin{figure}[htp]
\centering
\includegraphics[width=3.0in, height=1.5in]{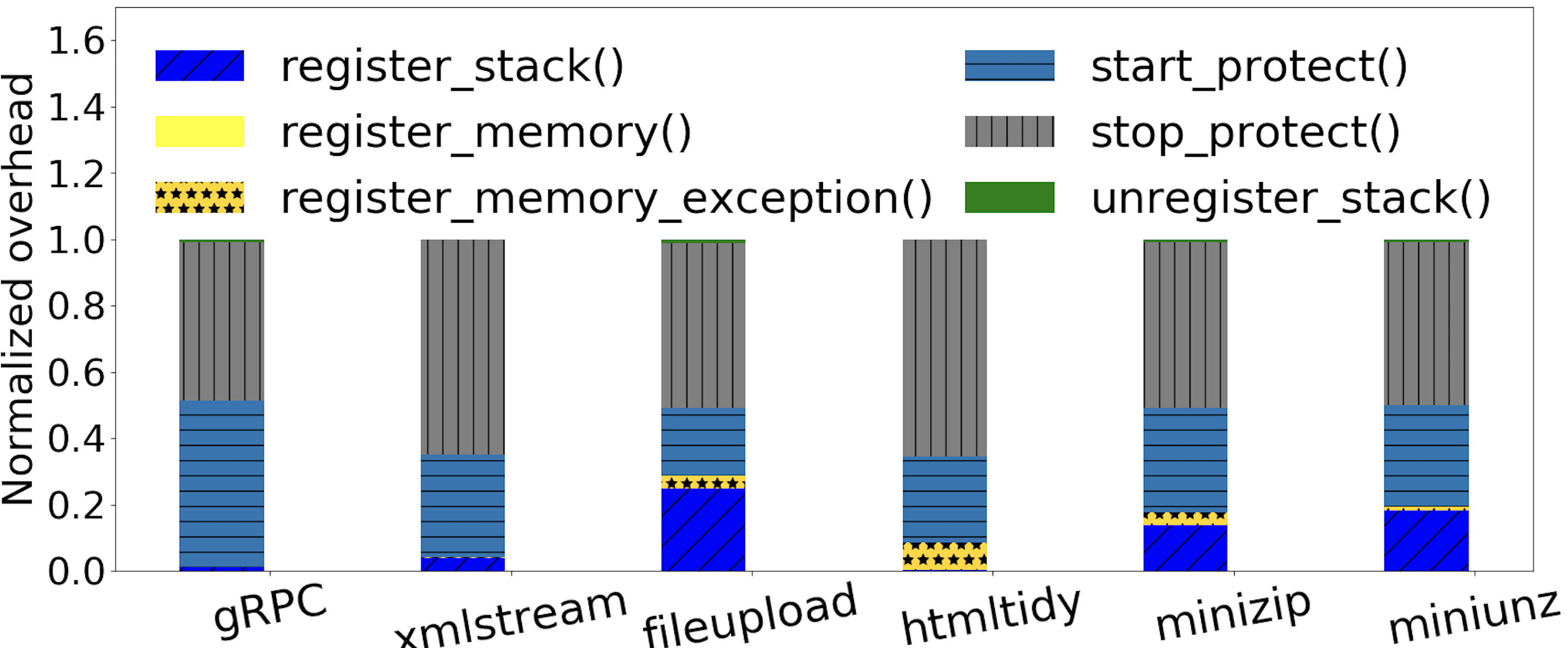}
\caption{Normalized overhead breakdown - per StackVault system call.}
\label{function_overhead_breakdown}
\end{figure}

We further divide the overhead of the StackVault system calls into each function, and the results are shown in Figure \ref{function_overhead_breakdown}. It can be observed that in most cases, \emph{start\_protect()} and \emph{stop\_protect()} incur the most overhead among all the StackVault specific system calls. These two system calls are the most sophisticated ones among all the six. Both of them need to copy data between the user stack and the kernel buffer, clear the user stack or free the kernel buffer, and refer to the memory areas registered by \emph{register\_memory()} and  \emph{register\_memory\_exception()} to make sure these areas are properly handled. Since a sensitive function can include multiple untrusted functions, \emph{start\_protect()} and \emph{stop\_protect()} can be invoked many times between a single pair of \emph{register\_stack()} and \emph{unregister\_stack()} calls. For each application, \emph{stop\_protect()} incurs more overhead than \emph{start\_protect()}. 
Also, \emph{unregister\_stack()} incurs slightly more overhead than  \emph{register\_stack()}. This is because \emph{unregister\_stack()} needs to clear the stack memory to make sure a sensitive function does not leave any sensitive data on the stack after it finishes running.

\subsection{Executable size overhead}
We compared the sizes of executable files between the Native case and the StackVault case. The results show that although the StackVault specific system calls are automatically inserted into the executable files when being compiled in the StackVault case, we see very little overhead in the executable file size. 
This is because compared with the number of instructions in the executable file, the number of inserted StackVault specific system calls is relatively small, which is negligible in this set of experiments.



Table \ref{executable_size} 
shows that the increased executable file size due to StackVault is within 0.5\%. Taking xmlstream for instance, with StackVault its executable file size increases 0.42\% from 13,416 bytes to 13,472 bytes. Since the number of inserted StackVault system calls is small compared to the size of application code, the larger the executable file is the less the relative overhead. This is demonstrated by gRPC, where the overhead is only 0.03\%. 

\begin{table}[]
\small
\centering
\caption{Comparison of executable file size(bytes) between the Native case and the StackVault case}
\label{executable_size}
\begin{tabular}{|c|c|c|c|ll}
\cline{1-4}
\textbf{Application} & \textbf{Native} & \textbf{StackVault} & \textbf{Overhead} &  &  \\ \cline{1-4}
gRPC                 & 37,281,600             & 37,282,936                & 0.03\%             &  &  \\ \cline{1-4}
xmlstream            & 13,416            & 13,472                & 0.42\%             &  &  \\ \cline{1-4}
fileupload           & 12,768            & 12,824                & 0.44\%             &  &  \\ \cline{1-4}
htmltidy             & 13,584            & 13,640                & 0.41\%             &  &  \\ \cline{1-4}
minizip              & 81,664           & 81,792               & 0.16\%             &  &  \\ \cline{1-4}
miniunz              & 77,200           & 77,336              & 0.18\%             &  &  \\ \cline{1-4}
\end{tabular}
\end{table}

\subsection{Compilation time overhead}

\begin{figure}[htp]
\centering
\includegraphics[width=3.0in, height=1.5in]{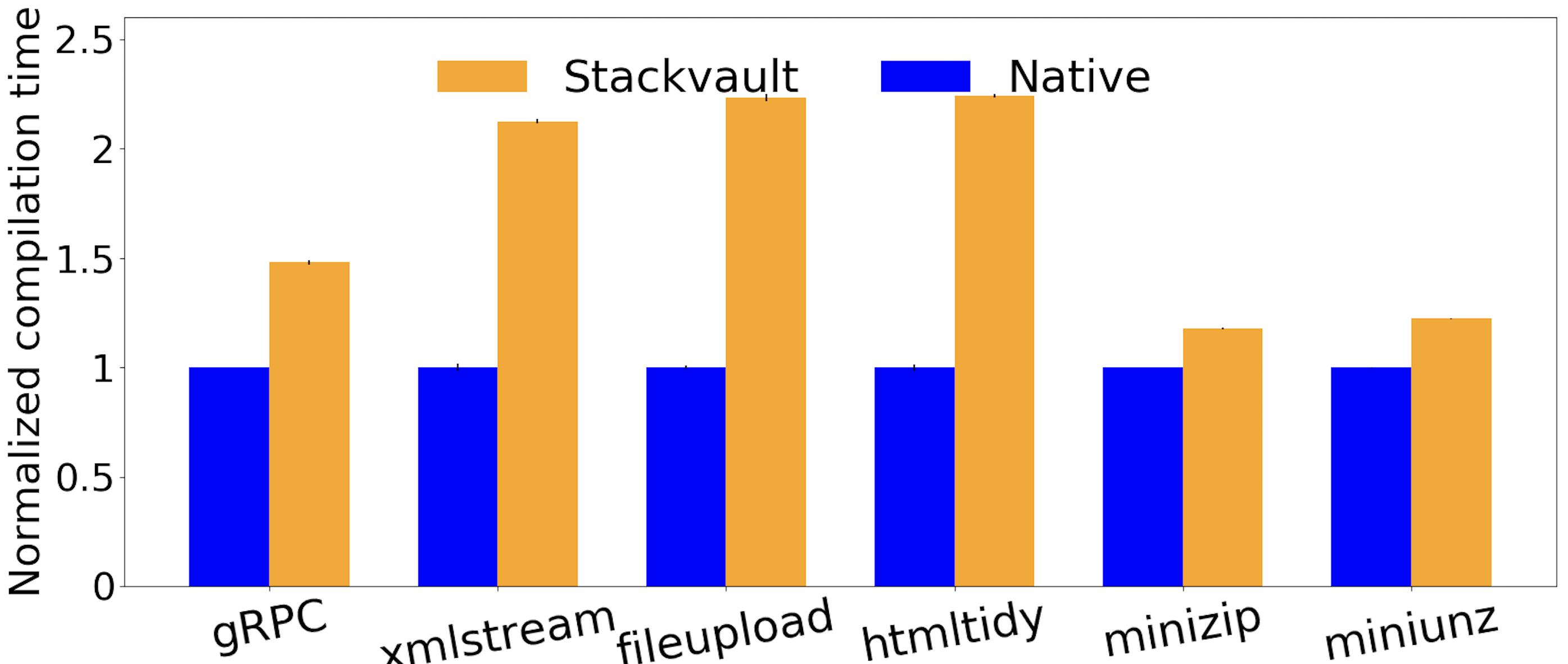}
\caption{Normalized compilation time comparison.}
\label{norm_compilation_time}
\end{figure}

Figure \ref{norm_compilation_time} compares the compilation time of each application between the Native and the StackVault cases. 
The compilation overhead is significantly high (about 1.24 times longer) for the smaller applications (xmlstream, fileupload, and htmltidy), 
but it becomes much smaller for larger application sizes, 
taking about 0.47, 0.17, and 0.22 times longer for gRPC, minizip, and miniunz, respectively. The StackVault system calls are currently implemented as a static library, and the compiler implementation uses link-time optimization to inline  these calls. This contributes significant compilation overhead, which can be optimized away by having the compiler directly generate the inline assembly instructions for each StackVault system call.

\section{Discussion}
The StackVault design handles recursive and nested functions for any ordering of sensitive and untrusted function calls.
It also handles untrusted functions that are part of dynamically loaded libraries.
If an untrusted function is invoked using a function pointer, 
the StackVault implementation is currently constrained to only handle them 
when the pointer can be resolved at compile time or link time.
However, in general it is possible to use runtime code specialization to handle all calls invoked using function pointers.


{\bf Protection for multithreaded applications.} In the presence of multiple threads, each thread has its own stack, but a thread can still access the stack-based data of another thread in the same process \cite{cross-thread}. 
A simple solution for the StackVault design to provide protection from untrusted function calls across threads is to use locks and priority queues to ensure that no \emph{new} sensitive function begins execution concurrently with the execution of an untrusted function.
This can be implemented with a guarantee of forward progress as long as untrusted functions do not have blocking dependencies on application code. This is a reasonable assumption for untrusted functions that are third-party library functions or reusable API calls.
This simple solution for multithreaded protection will add slight overhead to register\_stack, start\_protect, and stop\_protect calls. More sophisticated solutions that allow higher application concurrency are possible, but have to be balanced with the extra implementation overhead.




{\bf Vulnerabilities of other programming languages and Systems.}  
Although the current implementation of StackVault is based on C and C++, the mechanism introduced by StackVault can be applied to many other languages. Any programming model, language or runtime that relies on the assumption that "a function can access data in the memory of another function in the same process" is vulnerable to function-based data access attacks. A programming language is vulnerable to the function-based data access attacks if it satisfies one of the following criteria:

\begin{figure*}[]
	\centering
	\caption{Vulnerabilities of other programming languages and systems.}
	\includegraphics[width= 6.0in, height=1.0in]{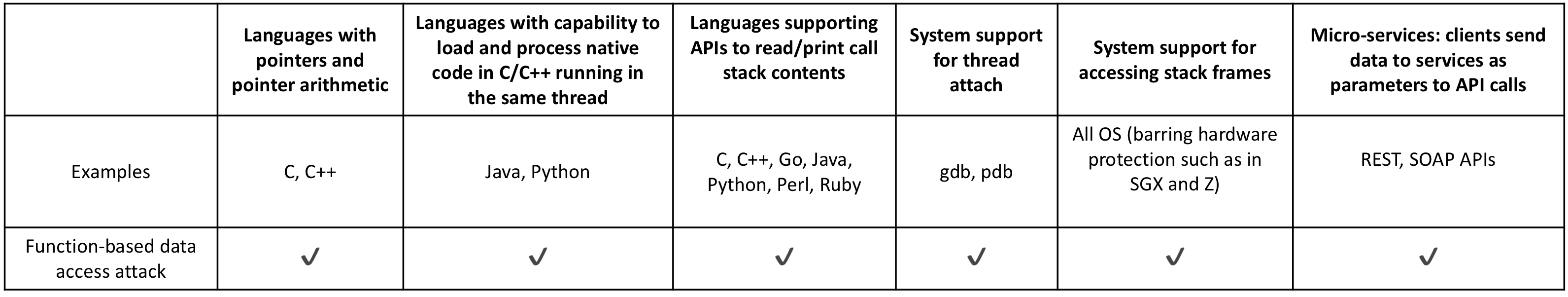}
	\label{fig:pls}
\end{figure*}

\begin{itemize}
    \item  It allows an untrusted function to directly or indirectly access the stack-based data of another function that has been called earlier.   
    \item It has a facility that allows functions written in C/C++ (languages that support pointers and pointer arithmetic) to be invoked, even if the language itself does not support pointers/pointer arithmetic (such as Java/Python). 
    \item It supports debugging APIs and tools to access/read call stack contents.
    \item It supports reflection that allows the program to introspect itself during execution. 
\end{itemize}
Figure~\ref{fig:pls} presents the types of programming languages and systems that are vulnerable to function-based data access attacks. 


     There are other systems and programming paradigms that can benefit from StackVault, e.g.~the micro-service paradigm.
     Micro-services support API calls, which are essentially function invocations in the form of remote procedure calls (over REST, SOAP). When such services receive parameters for the call, they use stack-frames to manage such data. The micro-services widely use third-party libraries in their implementation. Functions in those libraries (untrusted functions) when called from the API method implementation would allow these functions to access client-supplied data on stack-frames and/or in heaps pointed to by the stack-based variables.

\section{Related Work}

{\bf Stack based attacks.}\/ Due to the highly predictable layout of the stack memory, stack based attacks have existed for a long time, in which the most common one is buffer overflow \cite{chen2011linux}. For example, StackGuard \cite{cowan1998stackguard} proposes two techniques to overcome the buffer overflow vulnerability. One is putting a canary word right besides each return address on the stack, so that the modification of the return address can be detected by checking whether the canary word has been changed. This idea has also been incorporated in the GCC \cite{gcc} compiler. The other technique takes advantage of the debug registers to monitor the stack memory that stores the return address and triggers an exception once any return address has been rewritten. CRED \cite{ruwase2004practical} introduces a C range error detector, which allows programs to access out-of-bounds addresses that do not result in buffer overflows. Other stack protection approaches such as ASLR \cite{shacham2004effectiveness} and StackArmor \cite{chen2015stackarmor} use randomization to make it difficult for the attackers to guess where the target stack frame is. The shadow stack \cite{sinnadurai2008transparent, dang2015performance} was invented to protect return addresses on the stack from being tampered with. In this scheme, a shadow stack is maintained in parallel with the original stack, which is used to ensure the integrity of the address. 

Control-Flow Integrity (CFI) based approaches \cite{abadi2009control,criswell2014complete,zhang2013practical,abadi2005control} are also designed to protect stack-based buffer overflows. Such approaches first construct a Control-Flow Graph (CFG) using source code analysis, binary analysis, or execution profiling. Then, software execution will be dictated to follow the CFG, so that a compromised execution path caused by buffer overflow will be prevented. Also data leakages may happen via uninitialized reads. In this case, an attacker can get the stack data via reading uninitialized stack variables. \cite{milburn2017safeinit} and \cite{lu2016unisan} are two recent efforts that solve this problem by explicitly initializing each local variable after it is allocated on the stack. 

Strackx et al.~\cite{Strackx:2009:BMS:1519144.1519145} show how stack overreads can be used to carry out stack overflow attacks even in the presence of canaries and ASLR. The paper also talks about how stack overreads can also be used to override protections offered by memory obfuscation~\cite{bhatkar2003address} and instruction set randomization~\cite{kc2003countering}. Kundu and Bertino~\cite{placementnew} presented how placement new in C++ can be exploited for carrying out buffer overflow  as well as function pointer subterfuge attacks.


{\bf Memory data leakage.\/} There is a long line of research on preventing memory data leakage. For instance, Shreds \cite{chen2016shreds} protects the sensitive information in private memory by using the memory domain features in ARM CPU and \cite{wang2015between} explores the trust issues in multithreaded applications such as MemCached. \cite{halderman2009lest} finds that DRAM can retain its data for several seconds after it is powered off and removed from the motherboard and briefly discusses several solutions to these attacks, such as changing the architecture of the DRAM to make it lose state more quickly. SWIPE \cite{gondi2014minimizing, gondi2012swipe} takes advantage of static analysis to erase the sensitive data at the earliest time. \cite{chow2005shredding} presents a secure deallocation strategy to reduce the life cycle of the sensitive information in memory. Vanish \cite{geambasu2009vanish} aims at creating self-destruct data that can automatically vanish when it is no longer useful. All of these efforts try to reduce the probability of data leakage by reducing its lifetime. However, sensitive data can still be leaked within its lifetime.

{\bf Hardware protection.} 
SGX~\cite{costan2016intel} hardware protects a portion of an application and data processed from threats outside the SGX enclave.  SecureBlue++~\cite{boivie2012secureblue++} supports protection of a portion or the entire application or VM and the associated data  from threats outside the application runtime. In contrast, the threat we are addressing in this paper is inside-out, that is the data can be accessed and exfiltrated from inside the application to outside. 

In comparison, StackVault differs from the above mentioned works in that it identifies and reduces the vulnerabilities for stack-based attacks due to no isolation support among the functions within the same process. Moreover, existing memory isolation domains are at the process level, which are too coarse to prevent in-process function level illegal memory access.

\section{Conclusions and Future Work}
Untrusted third party libraries and malicious insider actors are becoming a significant threat leading to data leakage and exfiltration attacks. In this paper, we have described how function-based data access attacks can be enabled by malicious code in untrusted functions originating from such threats. We presented the StackVault system to protect stack-based sensitive data from being maliciously accessed by untrusted functions in the same process. StackVault introduces a set of programmer APIs, compiler extensions, and system calls  that can be used to protect sensitive data  during execution of untrusted functions and clear out the sensitive data from memory as soon as the sensitive function returns. StackVault uses a novel notion of unforgeable function identity in order to ensure that it will not be abused, subverted or spoofed by malicious functions. We have provided a security analysis and experimental evaluation of StackVault using popular real world applications such as gRPC. The results show that StackVault assures sufficient security guarantees while being highly efficient -- it incurs very low overhead on the execution time of applications. StackVault can be used to enhance compliance with GDPR and HIPAA by protecting sensitive personal and healthcare data.

In the future, we are planning to develop schemes to protect the contents of heaps from function-based data access attacks (heap contents that are  referred to by any stack variable are already protected by StackVault as presented in this paper).
Another future direction is to evaluate different security solutions for multi-threaded applications.

\bibliographystyle{plain}
\bibliography{stack-vault}

\begin{thebibliography}{10}

\bibitem{gdpr}
European general data privacy regulation.
\newblock
  \url{https://eur-lex.europa.eu/legal-content/EN/TXT/HTML/?uri=CELEX:32016R0679&from=EN}.
\newblock Accessed: 2017-07-05.

\bibitem{hipaa}
European general data privacy regulation.
\newblock \url{https://www.hhs.gov/hipaa/for-professionals/}.
\newblock Accessed: 2017-07-05.

\bibitem{gcc}
G{C}{C}, the gnu compiler collection.
\newblock \url{https://gcc.gnu.org/}.
\newblock Accessed: 2016-10-30.

\bibitem{cross-thread}
Cross-thread stack access.
\newblock
  \url{https://software.intel.com/en-us/inspector-user-guide-linux-cross-thread-stack-access},
  accessed Aug. 5, 2018.

\bibitem{fileupload}
Fileupload - upload to a file to a specific url using libcurl.
\newblock \url{https://curl.haxx.se/libcurl/c/fileupload.html}, accessed August
  2, 2018.

\bibitem{gRPC}
g{R}{P}{C} - a high performance, open-source universal rpc framework.
\newblock \url{https://grpc.io/}, accessed August 2, 2018.

\bibitem{htmltidy}
Htmltidy - download a document and use libtidy to parse the html.
\newblock \url{https://curl.haxx.se/libcurl/c/htmltidy.html}, accessed August
  2, 2018.

\bibitem{miniunz}
Miniunz - an open source file extraction tool.
\newblock
  \url{https://github.com/madler/zlib/blob/master/contrib/minizip/miniunz.c},
  accessed August 2, 2018.

\bibitem{minizip}
Minizip - an open source file compression tool.
\newblock
  \url{https://github.com/madler/zlib/blob/master/contrib/minizip/minizip.c},
  accessed August 2, 2018.

\bibitem{xmlstream}
Xmlstream - stream-parse a document using the streaming expat parser.
\newblock \url{https://curl.haxx.se/libcurl/c/xmlstream.html}, accessed August
  2, 2018.

\bibitem{teslaHack}
Elon musk is accusing a tesla employee of trying to sabotage the company.
\newblock
  \url{https://www.businessinsider.com/tesla-employee-engaged-in-sabotage-against-the-company-report-2018-6},
  accessed August 5, 2018.

\bibitem{githubHack}
Github hacked, millions of projects at risk of being modified or deleted.
\newblock
  \url{https://www.extremetech.com/computing/120981-github-hacked-millions-of-projects-at-risk-of-being-modified-or-deleted},
  accessed August 5, 2018.

\bibitem{equifax}
Equifax blames open-source software for its record-breaking security breach:
  Report.
\newblock
  \url{https://www.zdnet.com/article/equifax-blames-open-source-software-for-its-record-breaking-security-breach},
  accessed July 5, 2018.

\bibitem{hackedrepo}
Major open source code repository hacked for months.
\newblock
  \url{https://www.geek.com/news/major-open-source-code-repository-hacked-for-months-says-fsf-551344},
  accessed July 5, 2018.

\bibitem{abadi2005control}
Mart{\'\i}n Abadi, Mihai Budiu, Ulfar Erlingsson, and Jay Ligatti.
\newblock Control-flow integrity.
\newblock In {\em Proceedings of the 12th ACM conference on Computer and
  communications security}, pages 340--353. ACM, 2005.

\bibitem{abadi2009control}
Mart{\'\i}n Abadi, Mihai Budiu, {\'U}lfar Erlingsson, and Jay Ligatti.
\newblock Control-flow integrity principles, implementations, and applications.
\newblock {\em ACM Transactions on Information and System Security (TISSEC)},
  13(1):4, 2009.

\bibitem{bhatkar2003address}
Sandeep Bhatkar, Daniel~C DuVarney, and Ron Sekar.
\newblock Address obfuscation: An efficient approach to combat a broad range of
  memory error exploits.
\newblock In {\em USENIX Security Symposium}, volume~12, pages 291--301, 2003.

\bibitem{bishop2003computer}
Matt Bishop.
\newblock {\em Computer security: art and science}.
\newblock Addison-Wesley Professional, 2003.

\bibitem{boivie2012secureblue++}
Rick Boivie and Peter Williams.
\newblock Secureblue++: Cpu support for secure execution.
\newblock {\em IBM, IBM Research Division, RC25287 (WAT1205-070)}, pages 1--9,
  2012.

\bibitem{chen2011linux}
Haogang Chen, Yandong Mao, Xi~Wang, Dong Zhou, Nickolai Zeldovich, and M~Frans
  Kaashoek.
\newblock Linux kernel vulnerabilities: State-of-the-art defenses and open
  problems.
\newblock In {\em Proceedings of the Second Asia-Pacific Workshop on Systems},
  page~5. ACM, 2011.

\bibitem{chen2015stackarmor}
Xi~Chen, Asia Slowinska, Dennis Andriesse, Herbert Bos, and Cristiano
  Giuffrida.
\newblock Stackarmor: Comprehensive protection from stack-based memory error
  vulnerabilities for binaries.
\newblock In {\em NDSS}, 2015.

\bibitem{chen2016shreds}
Yaohui Chen, Sebassujeen Reymondjohnson, Zhichuang Sun, and Long Lu.
\newblock Shreds: Fine-grained execution units with private memory.
\newblock In {\em Security and Privacy (SP), 2016 IEEE Symposium on}, pages
  56--71. IEEE, 2016.

\bibitem{chow2005shredding}
Jim Chow, Ben Pfaff, Tal Garfinkel, and Mendel Rosenblum.
\newblock Shredding your garbage: Reducing data lifetime through secure
  deallocation.
\newblock In {\em USENIX Security}, pages 22--22, 2005.

\bibitem{costan2016intel}
Victor Costan and Srinivas Devadas.
\newblock Intel sgx explained.
\newblock {\em IACR Cryptology ePrint Archive}, 2016:86, 2016.

\bibitem{cowan1998stackguard}
Crispan Cowan, Calton Pu, Dave Maier, Jonathan Walpole, Peat Bakke, Steve
  Beattie, Aaron Grier, Perry Wagle, Qian Zhang, and Heather Hinton.
\newblock Stackguard: Automatic adaptive detection and prevention of
  buffer-overflow attacks.
\newblock In {\em Usenix Security}, volume~98, pages 63--78, 1998.

\bibitem{criswell2014complete}
J~Criswell, N~Dautenhahn, and VA~Kcofi.
\newblock Complete control-flow integrity for commodity operating system
  kernels.
\newblock In {\em 2014 IEEE Symposium on Security and Privacy (SP)}, pages
  292--307.

\bibitem{dang2015performance}
Thurston~HY Dang, Petros Maniatis, and David Wagner.
\newblock The performance cost of shadow stacks and stack canaries.
\newblock In {\em Proceedings of the 10th ACM Symposium on Information,
  Computer and Communications Security}, pages 555--566. ACM, 2015.

\bibitem{geambasu2009vanish}
Roxana Geambasu, Tadayoshi Kohno, Amit~A Levy, and Henry~M Levy.
\newblock Vanish: Increasing data privacy with self-destructing data.
\newblock In {\em USENIX Security Symposium}, pages 299--316, 2009.

\bibitem{gondi2012swipe}
Kalpana Gondi, Prithvi Bisht, Praveen Venkatachari, A~Prasad Sistla, and
  VN~Venkatakrishnan.
\newblock Swipe: eager erasure of sensitive data in large scale systems
  software.
\newblock In {\em Proceedings of the second ACM conference on Data and
  Application Security and Privacy}, pages 295--306. ACM, 2012.

\bibitem{gondi2014minimizing}
Kalpana Gondi, A~Prasad Sistla, and VN~Venkatakrishnan.
\newblock Minimizing lifetime of sensitive data in concurrent programs.
\newblock In {\em Proceedings of the 4th ACM conference on Data and application
  security and privacy}, pages 171--174. ACM, 2014.

\bibitem{halderman2009lest}
J~Alex Halderman, Seth~D Schoen, Nadia Heninger, William Clarkson, William
  Paul, Joseph~A Calandrino, Ariel~J Feldman, Jacob Appelbaum, and Edward~W
  Felten.
\newblock Lest we remember: cold-boot attacks on encryption keys.
\newblock {\em Communications of the ACM}, 52(5):91--98, 2009.

\bibitem{jang2014safedispatch}
Dongseok Jang, Zachary Tatlock, and Sorin Lerner.
\newblock Safedispatch: Securing c++ virtual calls from memory corruption
  attacks.
\newblock In {\em NDSS}, 2014.

\bibitem{kc2003countering}
Gaurav~S Kc, Angelos~D Keromytis, and Vassilis Prevelakis.
\newblock Countering code-injection attacks with instruction-set randomization.
\newblock In {\em Proceedings of the 10th ACM conference on Computer and
  communications security}, pages 272--280. ACM, 2003.

\bibitem{kharbutli2006comprehensively}
Mazen Kharbutli, Xiaowei Jiang, Yan Solihin, Guru Venkataramani, and Milos
  Prvulovic.
\newblock Comprehensively and efficiently protecting the heap.
\newblock {\em ACM Sigplan Notices}, 41(11):207--218, 2006.

\bibitem{placementnew}
A.~Kundu and E.~Bertino.
\newblock A new class of buffer overflow attacks.
\newblock In {\em 2011 31st International Conference on Distributed Computing
  Systems}, pages 730--739, June 2011.

\bibitem{lu2016unisan}
Kangjie Lu, Chengyu Song, Taesoo Kim, and Wenke Lee.
\newblock Unisan: Proactive kernel memory initialization to eliminate data
  leakages.
\newblock In {\em Proceedings of the 2016 ACM SIGSAC Conference on Computer and
  Communications Security}, pages 920--932. ACM, 2016.

\bibitem{luszcz2018apache}
Jeff Luszcz.
\newblock Apache struts 2: how technical and development gaps caused the
  equifax breach.
\newblock {\em Network Security}, 2018(1):5--8, 2018.

\bibitem{heartbleed}
Alyssa Milburn, Herbert Bos, and Cristiano Giuffrida.
\newblock Cve-2014-0160: The heartbleed vulnerability.
\newblock 2014.

\bibitem{milburn2017safeinit}
Alyssa Milburn, Herbert Bos, and Cristiano Giuffrida.
\newblock Safe{I}nit: Comprehensive and practical mitigation of uninitialized
  read vulnerabilities.
\newblock 2017.

\bibitem{nikiforakis2013heapsentry}
Nick Nikiforakis, Frank Piessens, and Wouter Joosen.
\newblock Heapsentry: kernel-assisted protection against heap overflows.
\newblock In {\em International Conference on Detection of Intrusions and
  Malware, and Vulnerability Assessment}, pages 177--196. Springer, 2013.

\bibitem{novark2010dieharder}
Gene Novark and Emery~D Berger.
\newblock Dieharder: securing the heap.
\newblock In {\em Proceedings of the 17th ACM conference on Computer and
  communications security}, pages 573--584. ACM, 2010.

\bibitem{pattabiraman2008samurai}
Karthik Pattabiraman, Vinod Grover, and Benjamin~G Zorn.
\newblock Samurai: protecting critical data in unsafe languages.
\newblock In {\em ACM SIGOPS Operating Systems Review}, volume~42, pages
  219--232. ACM, 2008.

\bibitem{pittenger2016know}
Mike Pittenger.
\newblock Know your open source code.
\newblock {\em Network Security}, 2016(5):11--15, 2016.

\bibitem{ruwase2004practical}
Olatunji Ruwase and Monica~S Lam.
\newblock A practical dynamic buffer overflow detector.
\newblock In {\em NDSS}, volume~4, pages 159--169, 2004.

\bibitem{shacham2004effectiveness}
Hovav Shacham, Matthew Page, Ben Pfaff, Eu-Jin Goh, Nagendra Modadugu, and Dan
  Boneh.
\newblock On the effectiveness of address-space randomization.
\newblock In {\em Proceedings of the 11th ACM conference on Computer and
  communications security}, pages 298--307. ACM, 2004.

\bibitem{Silvestro2017FreeGuardAF}
Sam Silvestro, Hongyu Liu, Corey Crosser, Zhiqiang Lin, and Tongping Liu.
\newblock Freeguard: A faster secure heap allocator.
\newblock In {\em CCS}, 2017.

\bibitem{sinnadurai2008transparent}
Saravanan Sinnadurai, Qin Zhao, and Weng fai Wong.
\newblock Transparent runtime shadow stack: Protection against malicious return
  address modifications, 2008.

\bibitem{Strackx:2009:BMS:1519144.1519145}
Raoul Strackx, Yves Younan, Pieter Philippaerts, Frank Piessens, Sven Lachmund,
  and Thomas Walter.
\newblock Breaking the memory secrecy assumption.
\newblock In {\em Proceedings of the Second European Workshop on System
  Security}, EUROSEC '09, pages 1--8, New York, NY, USA, 2009. ACM.

\bibitem{wang2015between}
Jun Wang, Xi~Xiong, and Peng Liu.
\newblock Between mutual trust and mutual distrust: practical fine-grained
  privilege separation in multithreaded applications.
\newblock In {\em 2015 USENIX Annual Technical Conference (USENIX ATC 15)},
  pages 361--373, 2015.

\bibitem{prevent-heartbleed}
David~A. Wheeler.
\newblock How to prevent the next heartbleed.

\bibitem{WOLFE201815}
Michael Wolfe, Seyong Lee, Jungwon Kim, Xiaonan Tian, Rengan Xu, Barbara
  Chapman, and Sunita Chandrasekaran.
\newblock The openacc data model: Preliminary study on its major challenges and
  implementations.
\newblock {\em Parallel Computing}, 78:15 -- 27, 2018.

\bibitem{zhang2013practical}
Chao Zhang, Tao Wei, Zhaofeng Chen, Lei Duan, Laszlo Szekeres, Stephen
  McCamant, Dawn Song, and Wei Zou.
\newblock Practical control flow integrity and randomization for binary
  executables.
\newblock In {\em Security and Privacy (SP), 2013 IEEE Symposium on}, pages
  559--573. IEEE, 2013.

\end{thebibliography}

\end{document}